# SELF-ORGANIZED GRAPHENE/GRAPHITE STRUCTURES OBTAINED DIRECTLY ON PAPER

A. R. Mailian[1*], G. Sh. Shmavonyan[2], M. R. Mailian[3]

[1] Institute for Informatics, 1 P. Sevak str., 0014, Yerevan, Armenia,
[2] State Engineering University of Armenia, Department of Microelectronics
and Biomedical Devices, 105 Teryan str., Yerevan, 0009, Armenia,
[3] LTX-Credence Armenia, 2 Adonts str., 0014, Yerevan, Armenia

**Abstract**

Considerable research has been carried out for synthesizing graphene and related materials by a variety of simple processes. We experimentally investigated the properties of graphite layers produced by an easy and non-conventional method of repeatedly rubbing conventional random stacked graphite bulk against insulating and semiconductor substrates. The patterned structure composed of rubbed-off and transferred layers exhibits properties of a solid-state material with through-thickness anisotropy of carrier mobility reaching $\sim 10^3$ cm$^2$/V·sec at the surface. The surface of the structure demonstrates quality of more ordered and optically oriented mono– or few layer graphene shaped by self-organization process due to friction. Enhanced photoconductivity originating from modification of continuous and linear valence and conduction bands caused by interaction between 4 graphene layers made possible obtaining Raman spectra at near infrared excitation wavelength of 976 nm.

**Introduction**

Since few layer graphene (FLG) is regarded as a material having favorable properties for device applications [1-3], interest in its preparation and physical properties remains strong. Recently graphene layers were produced from carbon nanotubes by rubbing on glass substrate [4]. It was found that during rubbing a bar of compacted carbon nanotubes against the surface of rough glass, the multilayer nanotubes are torn apart or peeled off radially leaving graphene layers on the substrate. The rubbed surface of the parent bar also was found to comprise grephene. Rubbing was used to obtain few-layer FLG sheets on flexible polymer substrates [5-7]. The conventional carbon bar was rubbed on sand papers and then the layer was rubbed onto plastic sheets, leaving a layer of FLG.

In all cases of mechanical exfoliation the trace or thin graphite structure or just the pencil drawn line was disbelieved to have attractive electronic properties of a solid material and because of that, perhaps, have been overlooked for investigation. Nonetheless, its micro-scale thickness made us believe that the behavior of electron gas in this system ought to change and we expected it should expose specific properties.

Previously, we reported on electronic properties of rubbed off graphite layers [8]. This article summarizes the pioneering experimental study of some basic properties of structures obtained from disordered graphite bulk by rubbing on different insulating substrates.

**Experiment**

Samples of graphite thin structures were prepared by series of sequential rubbing parent random stacked graphite bars against the surface of insulating or semiconductor substrates along the same track (see inset of Fig.1). Along with commercial graphite we commonly used bars made of pencil leads of B or HB hardness. All sorts of graphite (highly oriented pyrolytic graphite-HOPG, conventional graphite used in electrical engineering and chemistry, etc.) showed the same behavior. As a substrate we mainly used paper for convenience though, again, all types of insulating substrates showed identical results.

---

[*] Corresponding author. E-mail address: amailian@ipia.sci.am

In order to peel off layers, the parent graphite rod or bar was pressed normal to the substrate surface with the pressure of ≥1MPa, which is higher than the force required for cleaving off a flake with the thickness of several atomic layers from the bulk [9]. Actually drawing a line is a mechanical modification by combined and uninterrupted operation of cleaving→transferring→pressing of flakes. Therefore, we call the patterned layers or drawn lines *Combined Mechanical Modified (CMM)* layers and the structures containing multiple CMM layers - CMM structures. The advantage of our preparation method lies in the easy way to realize and it avoids contamination or surface doping avoiding the contact with any other material.

The electrical resistance of the structures was measured by a conventional two-probe method, using gold contacts pressed on the surface of the structures.

Raman spectra at 976 nm excitation light wavelength were obtained using a ThermoNXR FT-Raman Module and a Renishaw unit was used to obtain Raman spectra at 514 nm.

**Results and Discussion**

We found that the CMM structures behave much like a solid-state material with a number of interesting physical properties common to layered fine structures. First, the carrier transport is dimensionality dependent. Namely, the electrical resistance measured between contacts attached to the structure surface undergoes drastic decrease with the number of cleaved and transferred layers, i.e. with the thickness of the CMM structure (Fig.1a). Typically, the conductance became clearly detectable after a couple of rubbings. Take for example, samples presented in Fig. 1a. At the same pressure (~5 MPa), and with the same graphite rod or bar, layers on glass paper and ceramic substrates exhibited conductance at $2^{nd}$ rubbing or transfer, while layers on ZnO ceramic and plastic substrates at $3^{rd}$ rubbing.

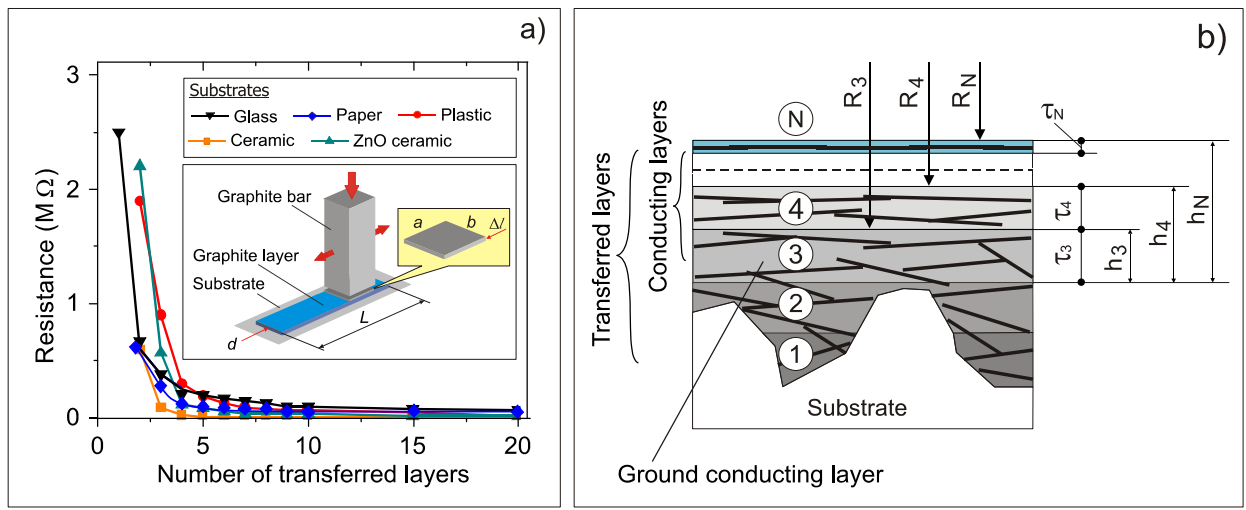

Fig.1. a) Typical dependence of the electrical resistance of CMM structures on the number layers transferred onto different insulating and semiconducting ceramic ZnO substrates at applied pressure of P ~ 5 MPa. Inset schematically illustrates the fabrication process of a CMM structure, where two red bold arrows indicate the direction of the movement of parent graphite bar during rubbing. b) Schematic presentation of packing of transferred layers in CMM structures. Vertical arrows show the surfaces from which the resistance was measured. Bold line segments represent the flakes.

During a few following rubbings the resistance dropped ~$10^3$ times, from several MΩ (first layer exhibiting conductance) to several hundred Ω, corresponding to low-resistance plateau (N>10). Note that the resistance of the structures never reached that of bulk. Such a behavior was the same for CMM structures obtained by a variety of graphite bars on different insulating and low-conductive semiconductor substrates and hence unmistakably reflects the intrinsic properties of graphite (Fig.1a).

Following the evolution of CMM structure under optical microscope, the pre-conduction structure was seen as a field of casual and rare spread spots or wormlike threads (in case of paper substrate) of several micrometer size (width of wormlike threads for paper substrates). This image should mean that initial rubbed off flakes merely embedded the pores of the substrate (layers 1 and 2, Fig.1b). In addition, the first 3-4 transferred layers of the CMM structure were highly disordered and contain mainly rippled areas and rare flat areas. The ground conducting layer and the layers shaped during further rubbing (N>3 in Fig.1) were shaped by a continuous chain of random stacked flakes bridging the graphite filled pores or graphite pools. Consequently, the conductance in these layers should be strongly limited because of extremely high boundary scattering between disordered and randomly stacked flakes.

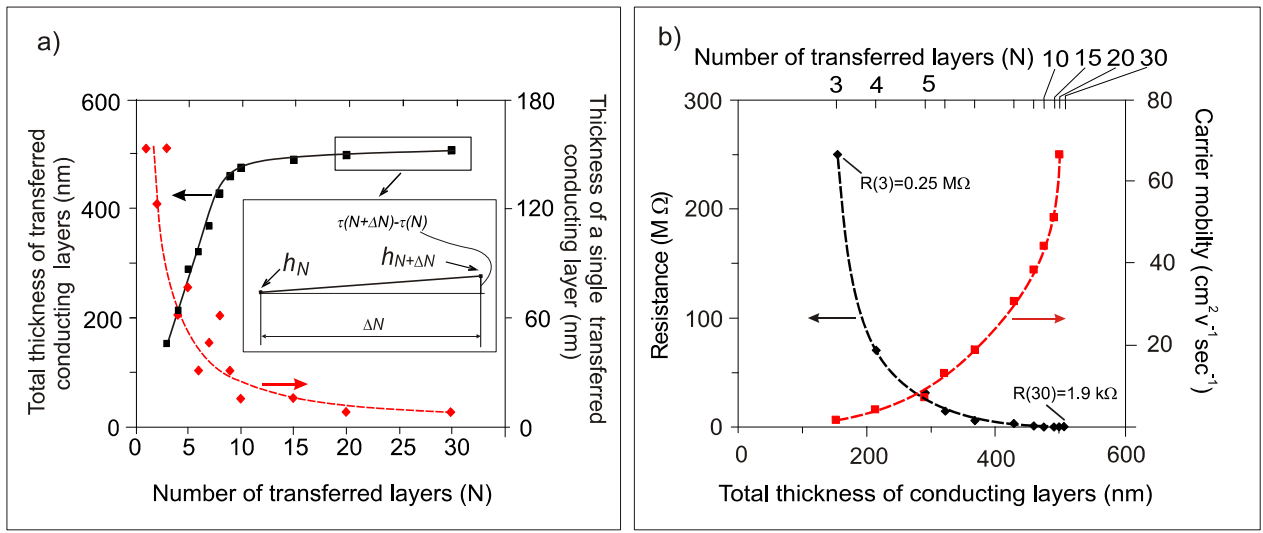

Fig. 2. a) Typical dependence of total thickness of the conducting layers in the CMM structure (black line) and the thickness of a single transferred layer (red line) on the number of the transferred layers on paper. The inset illustrates the calculation of thickness of one rubbed off layer. b) Typical dependence of resistance and carrier mobility of CMM structure on both total thickness of conducting layer (bottom scale) and the number of transferred layers (upper scale).

Hereafter we will examine the behavior of conducting layers shaped by both a single rubbing and layer-on-layer stacking. For convenience, the numeration of rubbing which is the same for transfer of cleaved off layers will be kept unchanged as shown in Fig.1b.

With increasing the number of the transferred layers (N =4-9) the disorder in layers decreases and the share of flat areas in the layers increases. Furthermore in the range of N =10-30, the surface of the CMM structure is covered mostly by flat areas. The thickness of $N^{th}$ individual transferred layer, $\tau_N$ was obtained by measuring the abraded volume of the parent graphite bar during rubbing (inset in Fig. 1a) [10]. The thickness of a single rubbed off and transferred layer is easily calculated by $\tau=a\Delta l/L$, where $L$ is the length of CMM layer, $\Delta l$ is the change in parent bar length during rubbing, $a$ is the width of the bar cross section and hence the width of CMM structure (inset in Fig.1). Then, the total thickness of the resultant conducting layer containing N transferred layers will be $h_N=\Sigma\tau_N$.

In reality, as the thickness of CMM structure increases, self-organization process due to the friction between the CMM surface layer and parent bar gradually strengthens and the growth rate of the structure slows down. Thickness vs. N dependence agrees well with this model - upper transferred individual conducting layers progressively get thinner (Fig. 2a). Also, the friction between the CMM structure and parent bar decreases significantly. Usually this means that the surface of the structure is ordered [4-7, 11-13].

The average thickness of $N^{th}$ layer in upper stratum of the CMM structure was estimated as $\tau_N=(h_{N+\Delta N} - h_N)/\Delta N$ (inset in Fig. 2a). For N=15÷30 corresponding to the linear plateau on resistance vs. *N* curve (Fig. 1a), $\tau_N$ ranged from ~ 1 to 6 nm depending on the surface roughness of a substrates and graphite hardness. Though obtained by macroscopic measurements, this nanoscale value is quite reasonable due to self-organization caused by shear stress between carbon layers in CMM structure and parent bar surface. Young's modulus which describes the shear strength of a material goes through a minimum when a graphite structure contains four graphene layers [14-16]. It means that four atomic layers make up a mechanically stable and stiff structure. The high value plateau on thickness vs. number of transferred layers is a sign of tightly packing and ordered stacking in CMM upper layers.

The carrier mobility is deduced from the measured electrical resistance of the CMM structure by $\mu_N = 1/en\rho_N = d/enbR_Nh_N$, where $\rho_N$ is the bulk resistance, *e* is the charge of an electron, *d* and *b* are the length and the width of a segment of the structure between measuring contacts, respectively, *n* - the bulk carrier density, and $R_N$ - the electrical resistance taken from the surface of CMM structure containing *N* transferred layers. For instance, for the sample presented in Fig. 2 ($\tau \approx 500$ nm) an enhancement in the mobility by a factor of 10s *(*Fig.2b) was obtained. But, apparent discrepancy between this value and sharp drop in the structure resistance hints on the assumption of isotropic CMM structure [8].

Next, we considered semilog plots of the inverse temperature dependence of inverse resistance for studied samples (Fig. 3b) extracted from resistance vs. temperature plots (Fig. 3a). We found that *ln(1/R)* vs. *T$^{-1}$* curves has linear slopes. Shallow electronic states responsible for linear slopes in *ln(1/R)* vs. *1/T* plot do not exist neither in single-layer nor in bulk of graphite. Rather they are band-edge states originating from modification of continuous and linear π and π∗ bands at K point of Brillouin zone due to interaction between basic layers [17]. Such T-behavior lends support for the model of through-thickness anisotropy. So, T-characteristics show that CMM structures demonstrate quasi two-dimensionality of carrier transport and provide clear evidence for layered and ordered stacking of CMM structures. [18-20].

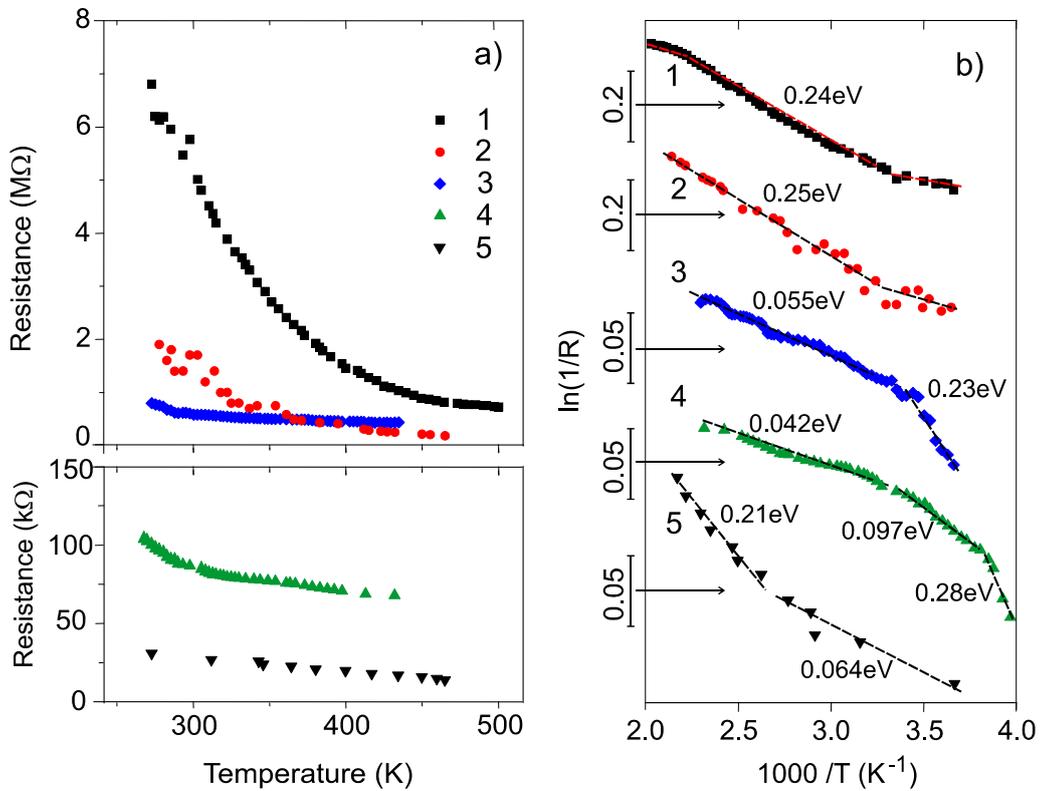

Fig. 3. a) Temperature dependence of electrical resistance of various CMM structures on paper. The structure resistance decreases by increasing the order of curves.

b) The same dependence in semilog plot of the inverse temperature. The bars on the left side present the scale. The figures in eV present the activation energies calculated for linear parts.

With regard to electrical measurements, we assume that because of strong screening the role of self-organized surface in carrier transport is crucial for the CMM structures. The inter-layer screening length to electric field in thin graphite is estimated to be 1.2 nm, which is about four graphene layer thick [21]. Thus, the surface layers of any ordered graphite structure is totally screened from the substrate due to huge anisotropy in in-plane and out-of-plane conductivity and sources of disorder originating from the substrate surface are reduced significantly reduces toward the surface [22-24]. We argue, therefore, that in our experiment mostly in-plane resistance of self-organized CMM top layer is measured from the contacts pressed on surface. Our observations and measurements too speak about independent behavior of transferred layers. Namely, the resistance of bottom transferred layers did not change with rubbing upper layers beginning from N=4.

We believe, therefore, it is four-layer flakes that are in majority in the transferred topmost layers and we would expect that they must greatly influence physical properties of the surface of a CMM structure. In support of the latter statement, the cut off distance for the Van der Waals force is assigned to be ~ 1 nm [25]. Electron dispersion contains prominent band edges above and under Fermi level at ~$1.61\gamma_1$ and ~$0.60\gamma_1$ (Fig. 4), where $\gamma_1$ is out-of-plane coupling parameter taken usually from 0.37 to 0.4 eV [26-33]. Corresponding energy gaps are ~1.27 eV and ~0.45 eV wide.

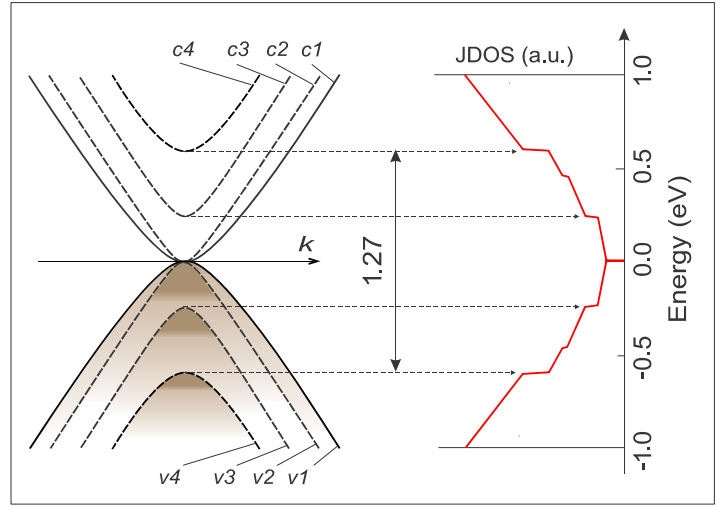

Fig. 4. Schematic presentation of the electron dispersion in four-layer graphene and joint density of states (JDOS) for Bernal stacking [29, 31].

We expected that the features of the layered CMM structure will be reflected in Raman spectra [34]. Raman spectrum obtained at 514 nm excitation wavelength looks like that of damaged graphene or nanostructured graphite with defects (Fig. 5) [35]. Note that $I_{2D}/I_G \approx 1$ indicates on the presence of FLG on the top of CMM. When incident laser light has the same energy ($E_{exc.}$) as the energy gap between two electronic states ($\Delta E$), i.e. when $E_{exc.} = \Delta E$ the resonant process increases the signal by a factor of ~$10^3$ in comparison to the intensity for a non-resonance Raman process [36]. Now, pursuing the assumption that flakes of four-layer graphene are in majority at the surface of CMM structure, we found 976 nm (1.27 eV) laser light suit well studying Raman scattering. On the other hand, the intensity of Raman scattering, depends on the wavelength of excitation light according to $\lambda^{-4}$ causing low signal/noise ratio at infrared. We believed that due to optical transition at 1.27 eV the resonant process will dominate over limitations in CMM structures.

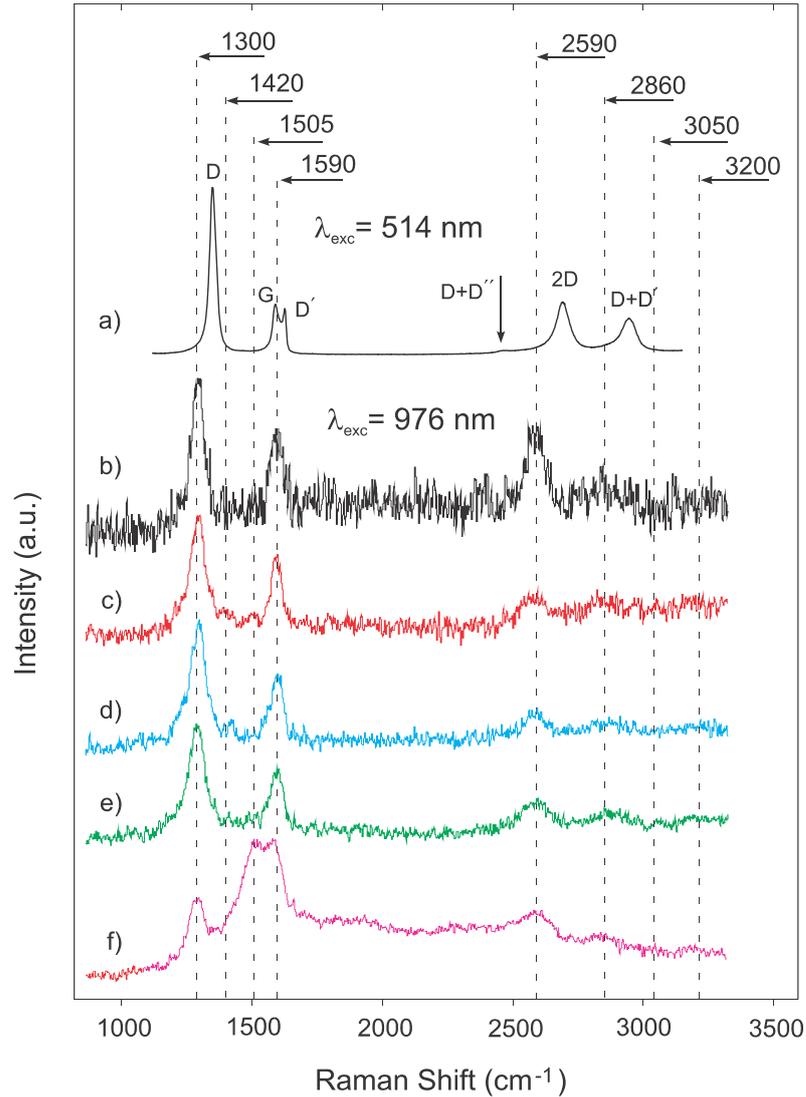

Fig. 5. Raman spectra of graphite and CMM samples at excitation light wavelength of 514 nm (a) and 976 nm (b-f). b) HOPG - as cut surface, c) HOPG - rubbed surface, d) random stacked conventional high conductive graphite – rubbed surface, e) CMM structure on random stacked graphite bulk, f) CMM structure on ceramic.

Raman spectra of CMM structures obtained at excitation wavelength of 976 nm are shown in Fig. 5 (spectra b-f). We probed samples in the following series: highly oriented pyrolytic (HOPG) – as-cut surface, HOPG – rubbed surface, random stacked conventional high conductive graphite – rubbed surface, CMM on random stacked graphite, CMM on ceramic substrate. All CMM samples were chosen to have lower resistance corresponding to plateau on resistance vs. transferred number dependence (N>20, Fig.1). It is important to note that as-cut surface of random stacked graphite did not reveal distinguishable Raman signal. This is a direct indication that limitations coming from longer wavelength of excitation light prevail over resonant enhancing of Raman signal. Meanwhile, self-organized surfaces or highly oriented graphite crystals were Raman active. We concluded therefore that strong optical transition at 1.27 eV present at the ordered and oriented topmost layer of a CMM structure enhanced signal/noise ratio in Raman spectra.

Judging from these results, the topmost layer of CMM structure is highly ordered. Our observations provide a hint on proving this assumption. We observed that the surface of CMM layer becomes lustrous when the resistance reaches the low value plateau shown in Fig. 1. We found also that the light (white and monochromatic) reflected from the surface of the CMM structure at an incident angle close to the structure surface is polarized parallel to the structure plane. This testifies to the existence of a self-organized, ordered and hence optically oriented crystal structure with c-axis parallel to the surface [37,38]. The intrinsic electrical screening between layers leads to picture out that bottom section of disordered layers serve as a substrate for upper ordered nanoscale layers i. e. the CMM structure is packed in Grapene/Graphite arrangement. In addition, our measurements of carrier transport in CMM structures show that the surface conductivity is possible to control by a gate voltage.

Thus, our assumption of dominant role four-layer flakes on the surface of rubbed off structure seems to be quite real, and since the surface is ordered stacked, it is natural to suppose that four-layer flakes shape a continuous lamina or stripes on the CMM surface.

The carrier mobility in top layer of CMM structure will be $\mu_N=1/en\rho_{xx}=a/enwR_N$, where $\rho_{xx}$ is in-plane resistivity at CMM structure surface. Assuming that $n \sim 10^{12}$ cm$^{-2}$ for conventional high conducting graphite, we obtained $\sim 3,300$ cm$^2$/V·sec. Now, this value matches well with the observed decrease in the electrical resistance of $\sim 10^3$ times.

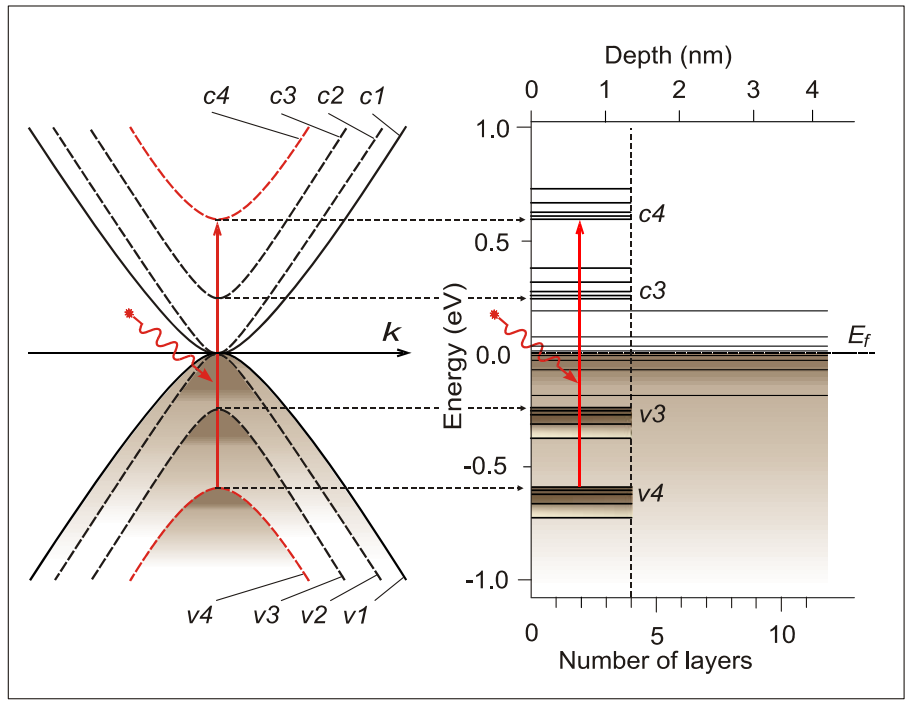

Fig. 6. Schematic representation of energy band diagram of the surface of CMM structure. The dark background represents continuous valence band at surface and in bulk.

In sum, our data give a solid ground to sketch out the energy band diagram of the surface of the CMM structures (Fig. 6). At the surface of CMM structure the electronic structure of hyperbolically dispersed continuous bands coexist with subbands with energy gaps of ~1.27 and ~0.45 eV. The first gap is between the top of forth valence and bottom of forth conduction bands (Ev4, Ec4), and the second one is between 3rd couple of bands (Ev3 and Ec3). Both type of electron dispersion (gapped and continuous) contribute in carrier transport and optical properties of CMM structure. Apparently, gapped electronic structure is responsible for prevailing selective conductivity which alters the properties of the surface of the CMM structure or pencil drawn line from semimetal toward those of semiconducting material.

**Conclusion**

In summary, it is found that graphite structures consisting of layers obtained by CMM exhibit peculiar physical properties such as:

- The CMM structure is packed in Graphene/Graphite arrangement due to friction-driven self-organization;
- The top layer of the structure containing mechanically transferred layers is ordered stacked and optically oriented and contains few-layer graphene;
- The CMM structures are electrically anisotropic through the thickness with the highest carrier mobility ~ 3,000 $cm^2/V \cdot sec$ at the top of the structure;
- Enhanced optical absorption due to the presence of photosensitive four-layer graphene on the surface of the CMM structure made it possible to obtain Raman spectra at the near infrared excitation light (976 nm). The spectra reflect the layered organization of CMM structures.

The overall results testify to CMM a graphitic system promising to obtain importance in obtaining selective carrier transport and high carrier mobility. The observed phenomenon is universal. It is observed regardless the substrate material and hence can find a successful application.

The suggested simple method allows one to obtain graphene with pencil directly on paper, as well as to design graphene-based electronic components and circuits on insulating substrates in particular on paper, which can enable flexible and cheap electronics.

**Acknowledgments**

The authors thank Dr. T. Kurtikyan for help with obtaining Raman spectra and M. Sharambeyan for helpful discussions.